%% file: top.tex
\documentclass[lnbip]{svmultln}
\usepackage{makeidx}  
\usepackage{eurosym}            
\usepackage{amssymb}
\usepackage{pifont}
\usepackage[capitalize]{cleveref}         
\usepackage{booktabs}         
\usepackage{graphicx}
\usepackage{xspace}
\usepackage{multirow}
\usepackage{longtable}
\usepackage{soul}
\usepackage{paralist}           

\usepackage[inline,finalnew]{trackchanges}
\addeditor{jn}

\input{definitions}

\begin{document}

\mainmatter              

\input{title}
\maketitle              

\input{abstract}

\section{Introduction}\label{sec:intro}
\input{intro}

\section{Background}\label{sec:background}
\input{background}

\section{Methods}\label{sec:method}
\input{method}

\section{Findings}\label{sec:findings}\label{sec:results}
\input{results}

\section{Discussion}\label{sec:discussion}
\input{discussion}

\section{Conclusions}\label{sec:conclusions}
\input{conclusions}


%
%
%
%
\bibliographystyle{splncs}
\bibliography{top}
%

%

%

\end{document}

%% file: definitions.tex
\newcommand{\Ocuco}{Ocuco\xspace}

\newcommand{\ProductOwner}{Product Owner\xspace}
\newcommand{\ProductOwners}{Product Owners\xspace}

\newcommand{\ScrumMaster}{Scrum Master\xspace}

\newcommand{\Developer}{Developer\xspace}

\newcommand{\ProjectManager}{Project Manager\xspace}

\newcommand{\SAFe}{SAFe\textsuperscript\textregistered\xspace}

%% file: title.tex
\title{Transition from Plan Driven to \SAFe: \\Periodic Team Self-Assessment}
\titlerunning{SAFe Team Self-Assessment}  
%
\author{Mohammad Abdur Razzak\inst{1} \and 
John Noll\inst{2} \and Ita Richardson\inst{1} \and Clodagh Nic Canna\inst{3} \and Sarah Beecham\inst{1}}
\authorrunning{Mohammad Abdur Razzak et al.}   
%
\tocauthor{Mohammad Abdur Razzak, John Noll, Ita Richardson, Clodagh Nic Canna, Sarah Beecham }
\institute{Lero, the Irish Software Research Centre, University of Limerick, Ireland\\
\email{{abdur.razzak, ita.richardson, sarah.beecham}@lero.ie} 
 \and
University of East London, University Way, London, E16 2RD, UK \\ 
 \email{j.noll@uel.ac.uk}
 \and
Ocuco Ltd, Blanchardstown Corporate Park, Dublin D15 N5DX, Ireland
 \email{clodagh.niccanna@ocuco.com } 
}

%% file: abstract.tex
\begin{abstract}
Context: How to adopt, scale and tailor agile methods depends on several factors such as the size of the organization, business goals, operative model, and needs. The Scaled Agile Framework (\SAFe) was developed to support organizations to scale agile practices across the enterprise. 

Problem: Early adopters of \SAFe tend to be large multi-national enterprises who report that the adoption of \SAFe has led to significant productivity and quality gains. However, little is known about whether these benefits translate to small to medium sized enterprises (SMEs).


Method: As part of a longitudinal study of an SME transitioning to SAFe we ask, \emph{to what extent are \SAFe practices adopted at the team level?} We targeted all team members and administrated a mixed method survey in February, 2017 and in July, 2017 to identify and evaluate the adoption rate of \SAFe practices.



Results: Initially in Quarter 1, teams were struggling with PI/Release health and Technical health throughout the organization as most of the teams were transitioning from plan-driven to \SAFe. But, during the transition period in Quarter 3, we observed discernible improvements in different areas of SAFe practice adoption.

Conclusion: 
 The observed improvement might be due to teams merely becoming more familiar with the practices over-time. However, management had also made some structural changes to the teams that may account for the change.
 


\keywords {SAFe, scrum, inter-team coordination, global software engineering, metrics, process assessment, software process improvement}
\end{abstract}

%% file: intro.tex

Software development is still driven by \emph{Infinite Diversity in Infinite Combinations} \cite{kuhrmann2015systematic}. As a consequence, practitioners ask themselves \emph{why they need to adopt} a practice, and \emph{how to scale} a practice. This leads to two challenges: first, recognising \emph{the purpose of a practice} and second, \emph{scaling practices}. Scaling agile continues to be a challenge in software development where the associated growth calls for strong coordination among teams as well as within the project \cite{abrahamsson2009lots, maples2009enterprise, turk2014limitations}. Scaling agile in globally distributed projects adds to the complexity  \cite{paasivaara2017adopting} since \emph{``Distance''} creates new challenges for successful scaling of agile practices.  

A number of frameworks have been proposed for scaling agile across the enterprise and the Scaled Agile Framework (\SAFe) is one of the most adopted of these models according to the Annual State of Agile Report \cite{11_versionone_report}. \SAFe has gained rapid attention amongst practitioners and is an important choice for organisations scaling agile development. Yet, the literature indicates that \SAFe is aimed at large-scale organizations. However, small-medium-enterprises (SMEs) are also interested in \SAFe as it provides an enterprise roadmap for adopting agile. 
 As the adoption of \SAFe increases, little research exists to identify how \SAFe is adopted in SMEs.  We conducted a study to measure the adoption of \SAFe recommended practices at the team level over time, in order to address the question \emph{How can the Scaled Agile Framework be implemented in an SME?}.

This paper is organised as follows: Section 2 provides a background to scaling agile frameworks, Section 3 presents the method we used in our empirical study, while Section 4 summarises our key findings and Section 5 discusses the implications of these results. Section 6 gives some conclusions to the study.



%% file: background.tex


\textbf{Agile Scaling Frameworks} Scaling agile covers the movement from a few agile teams to multiple agile development teams, where the number of teams can be in the hundreds \cite{ambler2008agile}. Scott Ambler \cite{ambler2008agile} pointed out several factors that need to be considered when scaling agile such as team size, geographical distribution, entrenched culture, system complexity, legacy systems, regulatory compliance, organizational distribution, governance and enterprise focus. In general, productivity and quality are the two main concerns of any organization when adopting a scaling agile paradigm. 

The choice of scaling agile framework which a company adopts or how the framework is tailored will depend on the organization's size or on ``what works'' based on their own business goals, operative model, and needs. The Agile Scaling Knowledgebase (ASK)\footnote{\path{http://www.agilescaling.org/home.html}} developed a matrix of different Agile frameworks namely \emph{Scrum-of-Scrum (SoS)\footnote{\url{https://www.agilealliance.org/glossary/scrum-of-scrums/}}}, \emph{Large Scale Scrum (LeSS)\footnote{\url{https://less.works}}}, \emph{Scaled Agile Framework (SAFe)\footnote{\url{http://www.scaledagileframework.com}}}, \emph{Disciplined Agile Delivery (DAD)\footnote{\url{http://www.disciplinedagiledelivery.com}}}, \emph{Spotify Model}\footnote{\url{http://blog.crisp.se/2012/11/14/henrikkniberg/scaling-agile-at-spotify}}, \emph{Nexus}\footnote{\url{https://www.scrum.org/resources/nexus-guide}}, and \emph{Scrum at Scale}\footnote{\url{https://www.scruminc.com/scrum-scale-case-modularity/}}. This matrix shows that \SAFe, launched in 2012 by Dean Leffingwell~\cite{Leffingwell_2015_Scaled} focuses on large enterprises and takes a scaled approach to agile adoption.

In comparison to (\SAFe), the other scaling agile frameworks (e.g;
SoS, LeSS, Nexus, Spotify) provide few artefacts, roles, and events in addition to Scrum. \SAFe provides more roles, events, artefacts and practices compared to other frameworks that enables \SAFe to scale on an organization level. The 11th Annual State of Agile report \cite{11_versionone_report} reported that, \SAFe is the most used scaling method used by 28\% respondents. In contrast, LeSS, DAD, and Nexus are reported to have a significantly lower take-up rate. 



\textbf{Scaled Agile Framework (\SAFe)} \SAFe is essentially a container for several existing agile approaches that is scalable and modular, and is primarily developed for organizing and managing agile practices in large enterprises. These qualities allow an organization to apply \SAFe in a way that suits their needs. Early adopters of \SAFe report that the application of the practices contained in this framework led to significant productivity and quality improvements \cite{laanti2014characteristics}. The literature also claims that \SAFe adoption is widespread including sectors such as manufacturing, software, and financial services \cite{laanti2014characteristics, paasivaara2017adopting, Pries_SAFeWay_2017, turetken2017assessing}. \SAFe 4.0 is organized into four layers: 1) \emph{Portfolio} -- Funding and coordinating programs, 2) \emph{Value Stream} -- Used when a single Agile Release Train (ART) cannot deliver the full solution, 3) \emph{Program} -- Contains 5-12 teams working towards a common goal, and 4) \emph{Team} -- Teams, which practice Scrum and/or eXtreme Programming and/or Kanban.



In \SAFe, all teams are part of the Agile Release Train (ART) and ARTs are the central construct of the program level. Teams are collectively responsible for defining, building and testing software in fixed-length iterations and releases. The team events (Backlog Refinement, Sprint Planning, Sprint Review) are  an integral part of \SAFe and help to reduce coordination overhead between teams. These teams typically consist of 7-9  members and teams operate on identical cadence and iteration lengths in order to provide better integration among teams \cite{turetken2017assessing}. But, adoption of only Scrum at the team level could lead to additional problems in task synchronization. To resolve this issue, \SAFe introduces the \emph{Release Planning} meeting to synchronize team tasks after every five iterations \cite{Leffingwell_2015_Scaled}. All teams on an ART are synchronized and integrated via common iterations that provide a valuable increment of new functionality. At the end of each iteration, the team perform a system demo for ART integration.



%% file: method.tex
%

\textbf{The Case Organization} The company we studied, \Ocuco, is a medium-sized Irish-based software company that develops practice and lab management software for the optical industry. \Ocuco employs approximately seventy staff members in its software development organization, including support and management staff. \Ocuco's annual sales approach \euro 20 million from customers across the British Isles, continental Europe, Scandinavia, North America, and China. 

\textbf{Data Collection} As part of a company-wide longitudinal study, we administered a SAFe self-assessment survey\footnote{\url{http://www.scaledagileframework.com/metrics/#T4}} to 70 team members in February, 2017 and in July, 2017. However, before the actual survey, two of the authors took a participant-observer role by sitting in on each team's Scrum ``ceremonies.'' One of us observed TeamA, daily, from January, 2016 to March, 2017, and TeamB, from May, 2017 to June, 2017; another of us observed TeamC, daily, from November, 2015 to July, 2016. We observed daily standups, sprint planning meetings, backlog grooming sessions, and sprint retrospectives. Due to the fact that the team members are distributed across Europe and North America, the observations were made by joining the video conference session for each ceremony. The observers also conducted semi-structured interviews with each member of the team he was observing, following an interview protocol \cite{Beecham_2017_Lean}.

\begin{table}[htbp]
\caption{List of participants.}
\label{tab:list-of-participants}
\centering
\scalebox{0.8}{
\begin{tabular}{lcc}
\hline
\textbf{Role} & {\textbf{Quarter 1}}  & {\textbf{Quarter 3}}      \\
&  (n=26) & (n=19) \\
\toprule
\ProjectManager (\ScrumMaster) & 9                    & 7                    \\ 
Developer                      & 9                    & 6                    \\
Quality Assurance              & 2                    & 3                    \\
Development Manager            & 1                    & --                   \\
Product Manager                & 2                    & --                   \\
Director of Eng.               & 1                    & --                   \\
Product Owner                  & 1                    & 3                    \\
Unclear                        & 1                    & --                   \\ 
\bottomrule
\end{tabular}}
\end{table}


The SAFe Self-Assessment survey comprises 25 questions that were sent to participants in an Excel Spreadsheet format. Each question has both a quantitative element (Likert scale), and an optional qualitative element (comment) that allowed participants to explain their ranking if needed. The Likert scale has six possible response options (ranging from `never'  to `always' as shown in Table \ref{tab:self-assessment-scale}) to measure the frequency of practice use according to each area (Product Ownership Health, PI/Release Health, Sprint Health, Team Health, and Technical Health).

In Quarter 1, we received 28 responses out of 70. Two responses were excluded as they were incomplete, resulting in a final set of 26, and in Quarter 3 we received 19 responses. The results represent a range of responses from seven roles. Table \ref{tab:list-of-participants} shows a breakdown of the roles of all 26 and 19 respondents (with one role unclear).

\begin{table}[htbp]
\caption{SAFe Team Self-Assessment scale.}
\label{tab:self-assessment-scale}
\centering
\scalebox{0.9}{
\begin{tabular}{lcccccc}
\hline
\toprule
\textbf{Value} & \textbf{0}   & \textbf{1}  & \textbf{2} & \textbf{3} & \textbf{4} & \textbf{5}   \\
\textbf{Meaning} & Never & Rarely & Occasionally & Often & Very Often & Always \\

\bottomrule
\end{tabular}}
\end{table}



\textbf{Data Analysis} To analyze the collected survey data, firstly, we extracted all qualitative and quantitative data. Secondly, we aggregated the 26 and then the 19 data points from Quarter 1 and Quarter 3 to get an overall view of all team members and to measure the frequency of practices used by teams according to each area (Product Ownership Health, PI/Release Health, Sprint Health, Team Health, and Technical Health) within the organization. Finally, we compared and contrasted across the two data sets to identify any changes over time.   


%% file: results.tex
%

\begin{figure}[htb]
\centering{
\includegraphics[width=0.7\columnwidth]{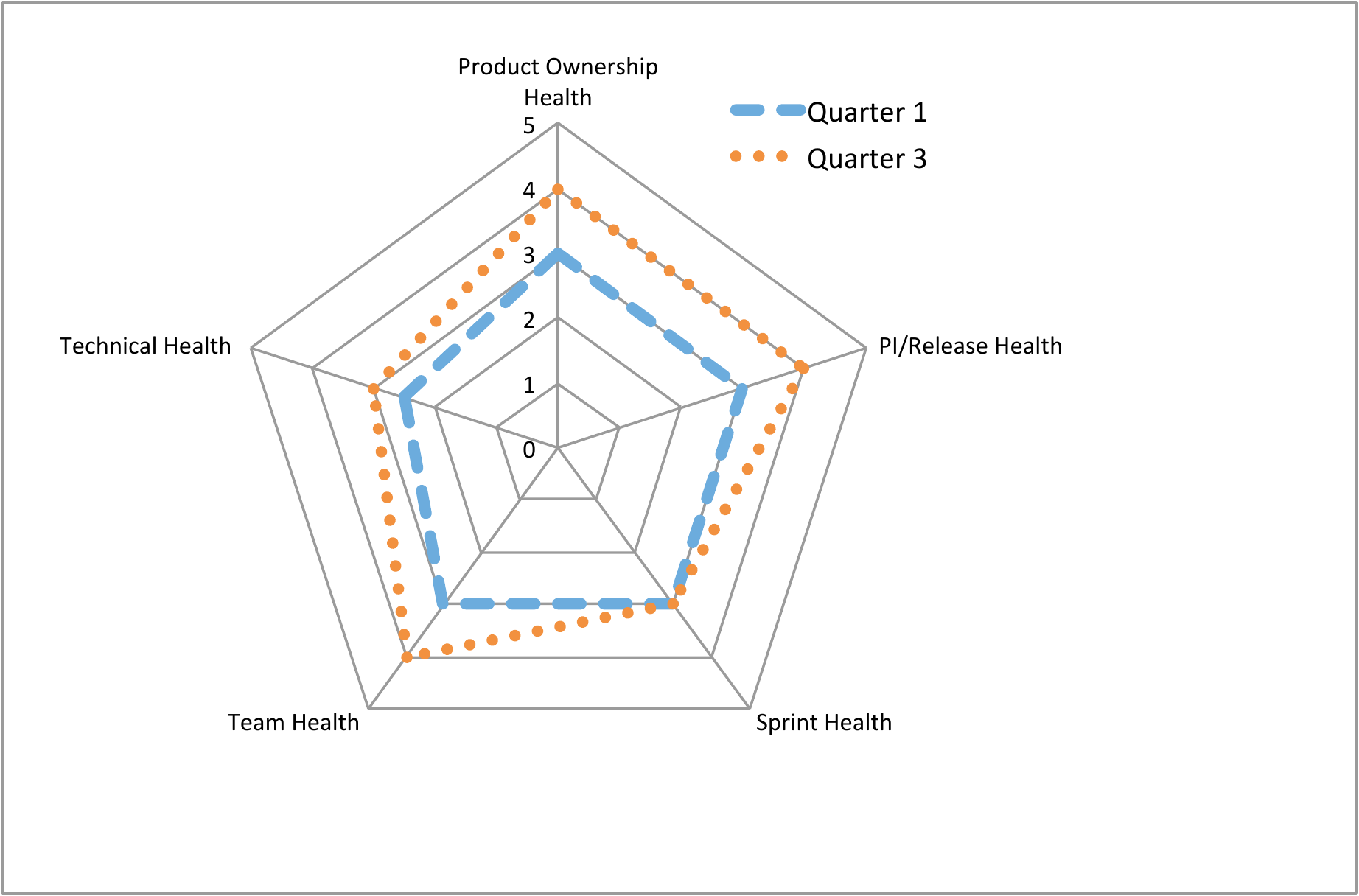}}
\caption{SAFe Team Self-Assessment (values: 0~-~Never, 1~-~Rarely, 2~-~Occasionally, 3~-~Often,  4~-~Very Often, 5~-~Always).}\label{fig:radar-chart}
\end{figure}

In this section, we present results of the qualitative and quantitative SAFe Team Self-Assessment. Figure \ref{fig:radar-chart} shows the median score across all participants. Of these, PI/Release health and Technical health were the most weak areas in Quarter 1 but responses to the repeated exercise undertaken in Quarter 3 indicates that there were marked improvements. 

\textbf{Product Ownership Health} Product Ownership means ensuring the success of the product, providing support, making a difficult decision when necessary, and considering the consequences of that decision \cite{Raithatha_2007}. In Scrum, the on-site customer role is fulfilled by a Product Owner, who represents the interests of the customer and end-users on a development team. Product Owners are responsible for communication between the customer and development teams \cite{hoda_impact_2011}. Product Owners also maintain the \emph{product backlog}, a list of user ``stories" that define requirements for the project. Table \ref{tab:product-ownership-health} shows the aggregated two stages result of Product ownership health at \Ocuco.

\begin{table}
\centering
\caption{Product ownership health.}
\label{tab:product-ownership-health}
\scalebox{0.8}{
\begin{tabular}{p{0.9\textwidth}ccc}
\toprule

\textbf{Question}& \textbf{Stage}  & \textbf{Median}$^1$ & \textbf{Mode}$^1$ \\
\midrule
\textbf{Q1.} Product Owner facilitates user story development, prioritization and negotiation  & \emph{Quarter 1}              & 3               & 3             \\ 
 & \emph{Quarter 3}         &     4         &   4          \\ 
\midrule
\textbf{Q2.} Product Owner collaborates proactively with Product Management and other stakeholders                     & \emph{Quarter 1}             & 4               & 4       \\ 
& \emph{Quarter 3}     &       4       &       4      \\ 
\midrule
\textbf{Q3.} User Stories are small, estimated, functional and vertical   & \emph{Quarter 1}   & 2.5 & 2,3 \\
& \emph{Quarter 3}      & 3              &   4          \\  
\midrule
\textbf{Q4.} Product owner facilitates development of acceptance criteria which are used in planning, review and story acceptance   & \emph{Quarter 1}                   & 3              & 4       \\ 
& \emph{Quarter 3}         &        4      &       4      \\ 
\midrule
\textbf{Q5.} Teams refine the backlog every sprint   & \emph{Quarter 1}                & 3               & 3       \\ 
& \emph{Quarter 3}      &     4         &   5          \\ 
\bottomrule
\multicolumn{3}{l}{$^1$ Values: 0 - Never, 1 - Rarely, 2 - Occasionally, 3 - Often,  4 - Very Often, 5 - Always.}
\end{tabular}}
\end{table}


Table \ref{tab:product-ownership-health} shows that there are three practices improved \emph{``Often''} to \emph{``Very Often''}, one \emph{``Occasionally''} to \emph{``Often''}, and one unchanged before and during operation.

 
According to quantitative data, at \Ocuco (Table \ref{tab:product-ownership-health}), the \ProductOwners use ``User Stories''  \emph{``Very Often''} but turning to the associated qualitative results, one of the \ProductOwners mentioned, 

\emph{\ldots{}We don't really use User Stories. We do a lot of prioritization and negotiation. We do a slightly more defined conversation/specification and communicate directly with developers.}

As a rationale for not using User Stories, a developer explained, 

\emph{\ldots{}This is a customer focused project. There is very little user story development in it. All we have are big long documents and specifications. However, they [\ProductOwners{}] did a good job in prioritizing and negotiating with the customer.}

This statement  results in our concluding that the \ProductOwner \emph{``Very Often''} facilitates prioritization, and negotiation (in Table \ref{tab:product-ownership-health}, Q1), and not user story development. 

But, on the other hand, a \ProjectManager who also acts as a \ScrumMaster said,

\emph{\ldots{}There is not a lot of negotiation going on for our team as the estimates are done in advance. Due to nature of contract we don't work with User Stories. We have deliverables that have been defined as part of the contract.} 

\textbf{PI/Release Health} 
In \SAFe, the Program Increment (PI) is the largest plan-do-check-adjust learning cycle that comprises PI planning, PI execution, the system demo, and the Inspect \& Adapt workshop respectively. Table \ref{tab:pi-release-health} shows the aggregated result of PI/Release health at \Ocuco.   

\begin{table}
\centering
\caption{PI/Release health.}
\label{tab:pi-release-health}
\scalebox{0.8}{
\begin{tabular}{p{0.9\textwidth}ccc}
\toprule
\textbf{Question} & \textbf{Stage}  & \textbf{Median}$^1$ & \textbf{Mode}$^1$ \\
\midrule
\textbf{Q1.}Team participates fully in Release Planning and Inspect and Adapt & \emph{Quarter 1}             & 3               & 3             \\ 
& \emph{Quarter 3}         &       4       &     4       \\ 
\midrule
\textbf{Q2.}Product backlog for the PI is itemized and prioritized  & \emph{Quarter 1}             & 3               & 3       \\ 
& \emph{Quarter 3}         &   4           &    4        \\ 
\midrule
\textbf{Q3.}Teams proactively interact with other teams on the train as necessary to resolve impediments & \emph{Quarter 1}             & 3               & 3       \\ 
& \emph{Quarter 3}       &  3             &      3      \\ 
\midrule
\textbf{Q4.}Team participates in System Demo every two weeks, illustrating real progress towards objectives  & \emph{Quarter 1}             & 3              & 3       \\ 
& \emph{Quarter 3}      &        4      &      5      \\ 
\midrule
\textbf{Q5.}Team reliably meet 80-100\% of non-stretch PI Objectives & \emph{Quarter 1}              & 3      & 3       \\ 
& \emph{Quarter 3}         &     3         &    3        \\ 
\bottomrule
\multicolumn{3}{l}{$^1$ Values: 0 - Never, 1 - Rarely, 2 - Occasionally, 3 - Often,  4 - Very Often, 5 - Always.}
\end{tabular}}
\end{table}


In Table \ref{tab:pi-release-health}, three practices improved from {\emph{``Often''} to \emph{``Very Often''}, and two practices were unchanged. In response to release planning, we received contradictory statements from two teams. The \ProjectManager said,  

\emph{\ldots{}We do not have a formal release planning, instead we plan continuously} 

But, a \ProductOwner said, 

\emph{\ldots{}All releases are planned. The whole team participates and know what is required for the version, and what can wait for the next in some cases.}

\textbf{Sprint Health} In Scrum, a \emph{sprint} is a set period of time during which specific work has to be done and made ready for review. During the planning meeting, the \ProductOwner and Agile team agree upon set of tasks needs to accomplish within a sprint based on the team bandwidth. Finally, the \ProductOwner defines the acceptance criteria for each assigned task to be completed at the end of a sprint. Table \ref{tab:sprint-health} shows the aggregated result of Sprint Health at \Ocuco.

\begin{table}
\centering
\caption{Sprint Health.}
\label{tab:sprint-health}
\scalebox{0.8}{
\begin{tabular}{p{0.9\textwidth}ccc}
\toprule
\textbf{Question} & \textbf{Stage}  & \textbf{Median}$^1$ & \textbf{Mode}$^1$ \\
\midrule
\textbf{Q1.}Team plans the sprint collaboratively, effectively and efficiently & \emph{Quarter 1}              & 4               & 4             \\ 
& \emph{Quarter 3}        &     4         &     3,5       \\ 
\midrule
\textbf{Q2.}Team always has clear sprint goals, in support of PI objectives, and commits to meeting them            & \emph{Quarter 1}                      & 3               & 3       \\ 
& \emph{Quarter 3}        &    4          &       3     \\ 
\midrule
\textbf{Q3.}Teams apply acceptance criteria and Definition of Done to story acceptance & \emph{Quarter 1}                                & 3               & 3      \\ 
& \emph{Quarter 3}      &      3        &       4     \\ 
\midrule
\textbf{Q4.}Team has a predictable, normalized velocity which is used for estimating and planning & \emph{Quarter 1}                              & 2.5              & 2       \\ 
& \emph{Quarter 3}     &     3         &     2,3,4       \\ 
\midrule
\textbf{Q5.}Team regularly delivers on their sprint goals  & \emph{Quarter 1}               & 3               & 3       \\ 
& \emph{Quarter 3}        &    3          &      3      \\ 
\bottomrule
\multicolumn{3}{l}{$^1$ Values: 0 - Never, 1 - Rarely, 2 - Occasionally, 3 - Often,  4 - Very Often, 5 - Always.}
\end{tabular}}
\end{table}


Table \ref{tab:sprint-health} shows , teams \emph{``Often''} calculate velocity to plan for the upcoming sprint. Additionaly, teams \emph{``Very Often''} plans the sprint collaboratively, effectively and efficiently, but one of the team members said, 


\emph{\ldots{}Sprints are not planned as such as we are at the tail end of the dev cycle. Almost all open tickets are go into the sprint.}


Though teams \emph{``Often''} calculate velocity to plan for the upcoming sprint, but due to lack of proper estimation, team cannot meet the sprint goals. 

\emph{\ldots{}We are often behind on doing the estimates, not taking the needed time or missing information enough to do a proper estimate}

A \ProjectManager identified, \emph{``Over commitment''} and QA \emph{``Speed''} are hindering the team in meeting the sprint goals. But, a \Developer said,
 
\emph{\ldots{}It's a bit up and down, sometimes we succeed. It is like it is become common to always introduce new `critical' issues into current sprint, instead of letting them wait for the next sprint planning.} 


\textbf{Team Health} There are three key roles defined in the Scrum development approach: the self-organizing development Scrum Team, the \ScrumMaster, and the \ProductOwner\cite{schwaber2002agile}. The \ScrumMaster is responsible for facilitating the development process, ensuring that the team uses the full range of appropriate agile values, practices and rules\cite{schwaber2002agile}. The \ScrumMaster conducts daily coordination meetings and removes any impediments that the team encounters\cite{schwaber2002agile}. Table \ref{tab:team-health} shows the aggregated result of Team health at \Ocuco.

\begin{table}
\centering
\caption{Team health.}
\label{tab:team-health}
\scalebox{0.8}{
\begin{tabular}{p{0.9\textwidth}ccc}
\toprule
\textbf{Question} & \textbf{Stage}  & \textbf{Median}$^1$ & \textbf{Mode}$^1$ \\
\midrule
\textbf{Q1.}Team members are self-organized, respect each other, help each other complete sprint goals, manage interdependencies and stay in-sync with each other & \emph{Quarter 1}     & 4   & 4         \\ 
& \emph{Quarter 3}         &  5             &    5        \\ 
\midrule
\textbf{Q2.}Scrum Master attends Scrum of Scrums and interacts with RTE as appropriate & \emph{Quarter 1}         & 3               & 4       \\ 
& \emph{Quarter 3}         &     3         &     5       \\ 
\midrule
\textbf{Q3.}Stories are iterated through the sprint with multiple define-build-test cycles (e.g. the sprint is not a waterfalled) & \emph{Quarter 1}             & 3               & 4      \\ 
& \emph{Quarter 3}       &      3        &      4      \\ 
\midrule
\textbf{Q4.}Team holds collaborative, effective and efficient planning and daily meetings where all members participate, status is given clearly, issues are raised, obstacles are removed and information exchanged  & \emph{Quarter 1}              & 4              & 4       \\ 
& \emph{Quarter 3}      &       5       &      5      \\ 
\midrule
\textbf{Q5.}Team holds a retrospective after each sprint and makes incremental changes to continually improve its performance     & \emph{Quarter 1}             & 3               & 4       \\ 
& \emph{Quarter 3}        &   4           &       5     \\ 
\bottomrule
\multicolumn{3}{l}{$^1$ Values: 0 - Never, 1 - Rarely, 2 - Occasionally, 3 - Often,  4 - Very Often, 5 - Always.}
\end{tabular}}
\end{table}

According to Table \ref{tab:team-health}, in \Ocuco teams \emph{``Always''} hold collaborative, effective and efficient planning meeting. Daily meetings are in place where all members participate, status is given clearly, issues are raised, obstacles are removed, and information exchanged among team members. Team members are self-organized, respect each other, \emph{``Always''} help each other to complete sprint goals, manage interdependencies, and stay in-sync with each other. 

Furthermore, team members are self-organized, respect each other, and help each other to complete sprint goals. A \ProductOwner states,  

\emph{\ldots{}Teams work well together and everyone is providing their part to making the best product. We just don't always agree on, which is good!}}


The Teams \emph{``Always''} hold collaborative, effective and efficient planning and daily meetings where all members, including remote team members, participate, status is given clearly, issues are raised, obstacles are removed and information exchanged with other team members. But, the teams rarely hold retrospectives after each sprint: 

\emph{\ldots{}I can only recall one retrospective during the last 2 years, it was done after a release and not after each sprint.}

\textbf{Technical Health} The Technical Health part of the survey helps a technology transformation team assess the current state of the technical maturity of a program/product line or organization. It can also be used later to have Agile teams assess their technical health and see if improvements have happened. The dimensions of the Technical Health part of the survey are: Continuous Delivery, Architecture, Technical Excellence, and Metrics. Table \ref{tab:technical-health} shows the aggregated result of Technical health at \Ocuco.

\begin{table}
\centering
\caption{Technical health.}
\label{tab:technical-health}
\scalebox{0.8}{
\begin{tabular}{p{0.9\textwidth}ccc}
\toprule
\textbf{Question} & \textbf{Stage} & \textbf{Median}$^1$ & \textbf{Mode}$^1$ \\
\midrule
\textbf{Q1.}Teams actively reduce technical debt in each sprint    & \emph{Quarter 1}              & 3               & 2             \\ 
& \emph{Quarter 3}         &   4           &        5    \\ 
\midrule
\textbf{Q2.}Team has clear guidance and understanding of intentional architecture guidance, but is free and flexible enough to allow emergent design to support optimal implementation  & \emph{Quarter 1}                             & 3               & 3,4       \\ 
& \emph{Quarter 3}          &    4          &      4      \\ 
\midrule
\textbf{Q3.}Automated acceptance tests and unit tests are part of story DoD    & \emph{Quarter 1}                             & 0               & 0      \\ 
& \emph{Quarter 3}         &      1        &      0      \\ 
\midrule
\textbf{Q4.}Refactoring is always underway        & \emph{Quarter 1}                        & 2.5              & 3       \\ 
& \emph{Quarter 3}        &       3       &  5          \\ 
\midrule
\textbf{Q5.}CI, build and test automation infrastructure is improving     & \emph{Quarter 1}                              & 2               & 0       \\ 
& \emph{Quarter 3}          &      3        &      0      \\ 
\bottomrule
\multicolumn{3}{l}{$^1$ Values: 0 - Never, 1 - Rarely, 2 - Occasionally, 3 - Often,  4 - Very Often, 5 - Always.}
\end{tabular}}
\end{table}

Interestingly, as Table \ref{tab:technical-health} shows, teams \emph{``Rarely''} adopt automated acceptance testing and unit testing as part of the story \emph{definition of done (DoD) \footnote{a list of criteria which must be met before a product increment ``often a user story'' is considered ``done''.}} \cite{schwaber2002agile}. 


Most of the teams at \Ocuco struggle with technical health, especially \emph{``test automation''} and  \emph{``refactoring''}. Throughout the organization none of the teams perform automated testing, but some teams are planning to adopt automatic test strategies. On the other hand, a Developer mentioned, there is no refactoring at all, because,


\emph{\ldots{}the customer keeps raising new requirements that contradicts with their previous requirements. Therefore, we kept adding new stuff while keeping the old one there because they might be worked on by a different developer, and we don't really know if they should just be removed/refactored. As a result, I can see quite a lot of functionality in the system that previously does the job but now it doesn't do anything and nobody is going to take them out as time goes by.}

%% file: discussion.tex



In software development, teams tailor their practices based on the metrics used to measure their system and evaluate their performance \cite{Leffingwell_2015_Scaled}. Agile teams continuously assess and improve their processes via a structured or periodic self-assessment as the first value of Agile Manifesto is to prefer \emph{``Individuals and interactions over processes and tools''}. By applying self-assessment, a software development team can understand its current process maturity, identify practices to improve, and practices that are missing. 

\textbf{Improving towards expectation?} The comparison shown in Table \ref{tab:comparison-round-1-2} (based on the Likert-scale results presented in Figure \ref{fig:radar-chart}) incorporating team improvement over time (5 months). In general, we observed a convincing improvement in four areas: Product Ownership Health, PI/Release Health, Team Health, and Technical Health but there was no discernible 
improvement in Sprint Health over the time. 

\begin{table}
\centering
\caption{Comparison.}
\label{tab:comparison-round-1-2}
\scalebox{0.8}{
\begin{tabular}{lccc}
\toprule
\textbf{Area} & \textbf{Quarter 1 (n=26)} & \textbf{Quarter 3 (n=19)} & \textbf{Improvement} \\
& February, 2017 & July, 2017 & 5 months \\
\midrule
Product Ownership Health  & 3 & 4 & +1           \\ 
\midrule
PI/Release Health & 3 & 4 & +1           \\ 
\midrule
Sprint Health & 3 & 3 & 0           \\ 
\midrule
Team Health & 3 & 4 & +1           \\ 
\midrule
Technical Health  & 2.5 & 3 & +0.5           \\ 

\bottomrule
\end{tabular}}
\end{table}

There appear to be several reasons for these observed improvements. As part of the company-wide longitudinal study three new dedicated \ProductOwners have been appointed as Management recognised that this is a full time job. One new \ProductOwner has prior knowledge about \SAFe. According to \Ocuco's Director of Development, ``we realized our Product Owners were being pulled in different directions by their multiple responsibilities, and as a result their teams were drifting away from the product roadmap. So we decided to hire additional staff so the Product Owners could focus solely on Product Ownership and keep the long-term product vision in-focus.'' This could a reason for the improvement Product ownership health as well as PI/Release health (Product backlog for the PI is \emph{``Very Often''} itemized and prioritized). 

There is some improvement in technical health moving from between \emph{``Occasionally''/``Often''} to \emph{``Often''}. According to \Ocuco's Director of Development, ``One of our new teams is adopting pure SAFe, to include \emph{automated test strategy} and \emph{continuous improvement} technique.'' That could be an another reason we are observing better results in Quarter 2 compared to Quarter 1. \Ocuco's Director of Development also mentioned, some teams are building their experience and learning over time. 

A major goal for \Ocuco  is to standardise their processes across all teams through transitioning to the \SAFe framework. They are starting to achieve this by tailoring \SAFe practices through modeling their ``as-is'' processes and identifying which practices need to be modified or added to achieve their target set of comprehensive ``to-be'' processes. Though \SAFe is primarily developed for organizing and managing agile practices in large enterprises it is clear that SME's are also interested in adopting \SAFe. However, \SAFe requires more roles, events, artefacts and practices compared to other frameworks to enable \SAFe to scale on an enterprise level. But, in SMEs it would be challenging if not impossible to adopt all the different ceremonies as well as fill all dedicated role such as Release Train Engineer (RTE). So, SME's need to consider which of the many ceremonies they want to adopt, and which roles they need to fill when adopting \SAFe.  They may also need to look at the various levels of Team Health and consider what level they want to reach that they feel is acceptable, when assessing how well they are doing against the  \SAFe self assessment survey results. 



%% file: conclusions.tex
%

In this study, we employed a mixed method approach to identifying and evaluating the adoption rate of agile practices as well as health levels of different process areas within a medium-sized Irish-based software company. Initially, we found that teams were struggling with PI/Release and Technical health throughout the organization as most of the teams were transitioning from plan-driven to \SAFe. But, during the transition over time, we observed a convincing improvement.

\SAFe provides more roles, events, artefacts and practices compared to other frameworks that aim to support organizations to scale on an enterprise level. But, in smaller organizations, adopting the many different ceremonies as well as dedicated roles may not be possible or necessary to meet their business goals. The results gained from the self-assessment at the Team level, may be satisfactory (there are only two practices in Quarter 3 that were reported as being used \emph{``Always''}, most reached a level of being used \emph{``Often''}). Therefore, as a result of our longitudinal study, we suggests that successful \SAFe implementation teams need to tailor the many \SAFe practices to understand the: 

 \textbf{Purpose of adopting a practice --\hspace{0.1cm}\emph{``Why''}} --\hspace{0.1cm} the Team needs to understand \emph{``why''} they need adopt agile practices.
 
\textbf{Implementation of a practice --\hspace{0.1cm}\emph{``How''}} --\hspace{0.1cm} the Team needs to learn \emph{``how''} to implement a practice to get the best out of it by tailoring \SAFe practices.

\textbf{Acknowledgments} We thank the members of TeamA, TeamB, and TeamC for their generous and thoughtful collaboration on this study, and for allowing us to study their software development efforts. This work was supported, in part, by Science Foundation Ireland grant 13/RC/2094  to Lero - the Irish Software Research Centre
(\url{www.lero.ie}).